\documentclass[aps,prl,twocolumn,twoside,floatfix,a4paper,superscriptaddress,showpacs]{revtex4}

\usepackage{bbm}
\usepackage{amsmath, amssymb}
\usepackage{color}
\usepackage{graphicx,epsfig}
\usepackage{dcolumn}
\usepackage{amsmath}
\usepackage{pstricks}
\usepackage{psfrag}

\newcommand{\ew}[1]{\langle #1 \rangle}
\newcommand{\partdiff}[2]{\frac{\partial #1}{\partial #2}}

\def\<{\langle}
\def\>{\rangle}

\begin{document}

\title{Validity of Landauer's principle in the quantum regime}

\author{Stefanie Hilt}
\affiliation{Department of Physics, University of Augsburg, 86135 Augsburg, Germany}
\author{Saroosh Shabbir}
\affiliation{Department of Physics \& Astronomy, University College London, London WC1E 6BT, UK}
\author{Janet Anders}
\affiliation{Department of Physics \& Astronomy, University College London, London WC1E 6BT, UK}
\author{Eric Lutz}
\affiliation{Department of Physics, University of Augsburg, 86135 Augsburg, Germany}

\date{\today}

\begin{abstract}
We demonstrate the validity of Landauer's erasure principle  in the strong coupling quantum regime by treating  the system-reservoir interaction  in a consistent way. We show that the initial coupling to the reservoir modifies both   energy and  entropy  of the system and provide explicit expressions for the latter in the case of a damped quantum harmonic oscillator.  These contributions are related to the Hamiltonian of mean force and dominate in the strong damping limit. They need therefore to be fully taken into account in any low-temperature thermodynamic analysis of quantum systems.
\end{abstract}
\maketitle

Information erasure is necessarily a dissipative process.  According to Landauer's principle  \cite{lan61},  erasure of one bit of information  requires a minimum  dissipation of heat of $kT \ln2$, where $T$ is the temperature and $k$ the Boltzmann constant.  As a result, the entropy of the environment increases by at least $k \ln2$. The erasure principle establishes a fundamental relationship between information theory and thermodynamics. As such, it has played a pivotal role in the resolution of Maxwell's demon paradox \cite{ben82}. The Landauer principle has been shown to hold for classical systems in the limit of  strong  \cite{lan61} and weak \cite{shi95,pie00}  interaction with the  reservoir, as well as for weakly damped quantum systems \cite{pie00}. However, its validity has recently been challenged in the  strongly coupled quantum regime. It has been  claimed that in the latter  the Clausius inequality,  $Q \leq  T\Delta S_{th}$, may break down \cite{all00,all02,hor05},  due to the entanglement of  system and reservoir  at very low temperatures (see Ref.~\cite{hil09} for a discussion of the precise role of entanglement).  This implies  that information may be erased, that  is,  entropy ($S_{th}$) decreased, while heat ($Q$)  is {\it absorbed}  \cite{all01,hor08}, in clear opposition to the  Landauer principle. This supposed violation has been reported in a growing number of recent books \cite{lef03,cap05,gem09} and reviews \cite{mar08}.
 Meanwhile, several key results of quantum information theory have been derived with the help of the erasure principle. Important examples include the Holevo bound on  accessible information \cite{ple99}, the no-cloning theorem \cite{ple01} and the upper bound on the efficiency of entanglement distillation \cite{ved00}. A failure of Landauer's erasure principle deep in the quantum domain would therefore have far-reaching consequences. 

In this paper, we resolve this quantum conundrum and show, by combining analytical and numerical analysis, that the Landauer principle {\it does} hold in the strongly coupled quantum regime. The resolution lies in a consistent thermodynamic treatment of the  coupling between system and reservoir. We therefore establish the  validity of the  erasure principle in classical and quantum physics, for arbitrary reservoir interaction strengths.

One of the basic assumptions of standard thermodynamics is that the system-reservoir coupling is negligibly small \cite{kub68}.  In this limit, a damped quantum system asymptotically relaxes  to a thermal Gibbs state \cite{ben81}. By contrast, for any finite interaction strength, the  quantum stationary state of the system  deviates from the Gibbs form, due to the noncommutation of position and momentum operators \cite{han05}. This leads to unexpected consequences.  Starting
from a microscopic  model for a strongly damped  harmonic oscillator, it has for instance been observed that, at zero temperature, the oscillator is  in an excited, mixed state and not in its pure ground state \cite{lin84,li95,jor04,for06}. Moreover, it has been shown that the Clausius inequality  is apparently violated at  low temperatures  during a quasistatic variation of the mass of the oscillator \cite{all00,all02,hor05,hil09,all01,hor08}. The Clausius inequality  asserts that, for a system initially in a thermal state, the change of entropy is always larger or equal than the amount of heat received by the system divided by the temperature \cite{kub68}; it is  regarded as a general formulation of the second law of thermodynamics. 
In the following, we resolve this paradoxical situation by considering, unlike in Refs.~\cite{all00,all02,hor05,hil09,all01,hor08},   the combined effect of  the mass variation {\it and} of the coupling to the reservoir. We explicitly show that the initial coupling modifies both  energy and  entropy (i.e. information content) of the oscillator. In the limit of strong coupling, these contributions  dominate those stemming from the variation of the mass. By properly taking into account the combined state transformation, initial reservoir coupling plus mass variation, we are  able to demonstrate the general validity of the Clausius inequality, and hence of the Landauer principle, in the  strongly coupled regime.

{\it Microscopic system-reservoir model}. Following Refs. \cite{all00,all02,hor05,hil09,all01,hor08}, we base our study of the Clausius inequality in the quantum domain on the standard model for quantum dissipation \cite{wei99}. The latter consists of a harmonic oscillator linearly coupled to a bath of \mbox{harmonic oscillators:}
\begin{align}\label{1}
	H = H_S + \sum_{j=1}^N\left[\frac{p_j^2}{2 m_j}	
              +\frac{m_j\omega_j^2}{2} \left(x_j-\frac{C_j q}{m_j\omega_j^2}\right)^2\right],
\end{align}
where  the $C_j$'s are  coupling constants. The Hamiltonian of the system  is $H_S=p^2/(2M)+ M \omega^2 q^2/2$ with the usual notation. The reservoir  is  characterized by the Ohmic spectral density, $J(\nu)= \pi/2\sum_j C_j^2/(m_j \omega_j)\,\delta(\nu-\omega_j)= \eta \nu \omega_D^2/(\nu^2+\omega_D^2)$, with damping coefficient $\eta$ and Debye cutoff frequency $\omega_D$ \cite{wei99}. System and  bath are  supposed to be initially decoupled and each in equilibrium at the same temperature $T$. The total density operator is thus  $\rho(0)=\rho_S(0) \otimes \rho_B(0)$ with $\rho_S(0) = \exp(-\beta H_S)/\mbox{Tr}_S \exp(-\beta H_S)$ and $\rho_B(0) = \exp(-\beta H_B)/\mbox{Tr}_B \exp(-\beta H_B)$; the bath Hamiltonian is $H_B= \sum_j p_j^2/(2 m_j)+m_j\omega_j^2 x_j^2/2$ and $\beta=(kT)^{-1}$. 

While the whole system equilibrates to the Gibbs state $\rho = \exp(-\beta H)/\mbox{Tr} \exp(-\beta H)$, the reduced stationary phase space distribution of the damped oscillator is non-Gibbsian; it is given by a Gaussian with variances \cite{gra84},
\begin{align}\label{2}
	\< q^2\> (\eta, M) &= \frac{\hbar}{M\pi} \sum_{i=1}^3 \left[\frac{(\lambda_i-\omega_D) \, \psi \left(1+\frac{\beta\hbar\lambda_i}{2\pi}\right)}{(\lambda_{i+1}-\lambda_i)(\lambda_{i-1}-\lambda_i)}\right]  \notag\\ &+ \frac{1}{M \beta \omega^2} \ ,\\
\label{2}	\< p^2\> (\eta, M) &=\frac{\hbar\eta\omega_D}{\pi} \sum_{i=1}^3 \left[\frac{\lambda_i \, \psi \left(1+\frac{\beta\hbar\lambda_i}{2\pi}\right)}{(\lambda_{i+1}-\lambda_i)(\lambda_{i-1}-\lambda_i)}\right] \notag\\
	&+ M^2\omega^2\ew{q^2}  (\eta, M)\ . 
\end{align}
The parameters $\lambda_i(\eta, M)$ are here the characteristic frequencies of the damped harmonic  oscillator and $\psi$ denotes the digamma function. To facilitate the following discussion, we have explicitly indicated the dependence on the coupling constant $\eta$ and on the mass $M$ in the above equations.  Due to the finite coupling to the reservoir, the variances are squeezed, $M\omega^2 \<q^2\> < \<p^2\>/M$, and the stationary state is hence non--thermal. For an isolated quantum oscillator ($\eta=0$), Eqs.~(2) and (3) reduce to their known thermodynamic expressions:  $\< q^2\>(0,M) = \hbar/(2M \omega) \coth(\beta \hbar \omega/2)$ and $\< p^2\>(0,M) = \hbar M \omega/2 \coth( \beta \hbar \omega/2)$.  

\begin{figure}
\begin{center}
 \includegraphics[width=.49\textwidth]{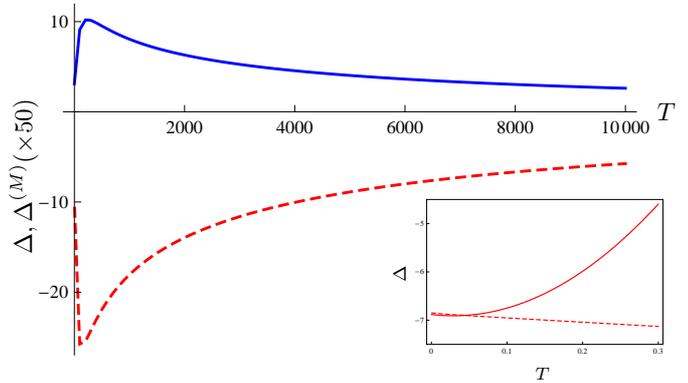} 
\end{center}
\caption{(Color online) Temperature dependence of the quantities $\Delta^{(M)} =  Q^{(M)} - k T \Delta S^{(M)}$ (blue solid) for the mass variation alone and $\Delta =  Q - k T \Delta S$ (red dashed) for the combined transformation mass variation plus initial coupling. The Clausius inequality is obeyed in the latter, while it appears violated in the former. A comparison between the exact (red dashed)  and the low-temperature approximation (red solid), Eq.~\eqref{14},   of  $\Delta$ is shown in the inset. Parameters are $\omega=1.2, M_0=1.1, M_1=1.11, \eta=40 \omega, \omega_D=25  \eta $ and $\hbar=k=1$.}\label{f1}
\end{figure}

{\it Entropy and heat for mass variation.} Suppose the oscillator undergoes a quasistatic mass variation from $M_0$ to $M_1$, as discussed in  Refs.~\cite{all00,all02,hor05,hil09,all01,hor08}. The internal energy of the system is defined as the stationary expectation value of its energy, $U=\<H_S\>=\<p^2\>/(2 M)+ M\omega^2 \<q\>^2/2$. As a result, the oscillator  exchanges an amount  of heat with the external reservoir given by \cite{all00,all02,hor05,hil09,all01,hor08},
\begin{align}\label{4}
	 Q^{(M)}&=\int_{M_0}^{M_1}\left(\frac{1}{2 M}\partdiff{\ew{{p}^2}}{M}+\frac{M\omega^2}{2}\partdiff{\ew{{q}^2}}{M}\right) dM\ .
\end{align}
At the same time, the von Neumann entropy of the quantum oscillator changes by $\Delta S^{(M)}= S(\eta, M_1) -  S(\eta, M_0)$, where $S= -\mbox{Tr} \rho_S \ln \rho_S$  can be expressed as \cite{all00,all02,hor05,hil09,all01,hor08},
\begin{align}\label{4}
	S= \left(v+ \frac{1}{2} \right)	\ln \left(v+ \frac{1}{2} \right) - \left(v- \frac{1}{2} \right) \ln \left(v- \frac{1}{2} \right) \ .	
\end{align}
Here $\rho_S=\mbox{Tr}_B \rho$ is the reduced density operator of the oscillator and $v=v(\eta, M) = \sqrt{\<q^2\>\<p^2\>}/\hbar$  the phase space volume. The temperature dependence of  $\Delta^{(M)} =  Q^{(M)} - k T \Delta S^{(M)}$ is shown in Fig.~\ref{f1} (the thermodynamic entropy is given by $S_{th} = k S$). We observe that  $\Delta^{(M)}$ is positive at very low temperatures, in apparent violation of the Clausius inequality which would require that  $\Delta^{(M)}\leq0 $  when $\Delta M = M_1-M_0\geq0$. This result should not surprise, as Clausius' inequality assumes that the system is initially in a thermal state \cite{kub68}, which is not the case in the strongly coupled regime.

Analytic expressions for  $Q^{(M)}$, $\Delta S^{(M)}$ and $\Delta^{(M)}$ can be derived close to zero temperature. A lowest order expansion of the variances  (2) and (3) in the limits $\omega_D\gg \eta/M\gg \omega$, yields \cite{gra84,wei99},
\begin{align}
	\ew{q^2}(\eta,M)&=\frac{2 \hbar}{\pi\eta}\ln\frac{\eta}{M\omega}
			+\frac{\pi\eta}{3\hbar M^2 \omega^4} \, (kT)^2 + \mathcal{O}(T^3)\ ,\\
	\ew{p^2}(\eta,M)&=\frac{\hbar\eta}{\pi}\ln\frac{\omega_D M}{\eta} + \mathcal{O}(T^3)\ .
\end{align}
With the help of the above expressions, we  obtain,
\begin{align}
	 Q^{(M)} &= (b_0 -b_1) \, \frac{\hbar \omega}{2 \pi} \, \left[1- \frac{\pi^2}{6 a^2}\right], \label{8}\\
\Delta S^{(M)} &=\frac{1}{2} \left[ \ln \frac{\ln c_1}{\ln c_0} - \ln \frac{\ln b_0}{\ln b_1}  - \frac{\pi^2 }{24 a^2}\left(\frac{b_0^2}{\ln b_0} - \frac{b_1^2}{\ln b_1}\right)\right], \label{9}
\end{align}
where we have defined the dimensionless parameters, $a = \beta \hbar \omega/2$, $b_i = \eta /(M_i \omega)$ and $c_i = M_i \omega_D/\eta$. In the low temperature, strong coupling regime, $Q^{(M)} $ is always positive, that is, heat is absorbed by the system, while information can be erased, $\Delta S^{(M)} <0$.  By further combining Eqs.~\eqref{8} and \eqref{9}, we  arrive at,
\begin{align}
	\Delta^{(M)} 	&= \frac{\hbar \omega (b_0 - b_1) }{2 \pi} \, \times \\ 
			&\left[1 - \frac{\pi}{(b_0-b_1) 2a} \, \left[ \ln \frac{\ln c_1}{\ln c_0} - \ln \frac{\ln b_0}{\ln b_1} \right] - \frac{\pi^2}{6 a^2}\right] \ . \notag
\end{align}
This  quantity is positive when $b_0>b_1 (\gg1$) and $a\gg1$.

{\it Entropy and heat for  coupling to the reservoir.} We  turn to the evaluation of entropy and heat resulting from  the initial coupling of the isolated  oscillator to the reservoir. In contrast to the previous case, the initial state of the system is now thermal. When the coupling coefficient is quasistatically increased from $0$ to $\eta$, the internal energy  changes by  $\Delta U^{(C)} = U(\eta, M_0) -  U(0, M_0)$. The corresponding heat can then be determined via the first law, $Q^{(C)} = \Delta U^{(C)}  - W^{(C)}$. We note that for a quasistatic transformation, the work required to couple the oscillator to the reservoir is given by the free energy difference $W^{(C)}=\Delta F^{(C)} = F(\eta, M_0) -  F (0, M_0)$ \cite{for06}. The latter can be evaluated using the general form  of the  free energy of a quantum damped oscillator \cite{gra84,wei99},
\begin{align} \label{11}
	\beta F 	&= \ln \Gamma\left(\frac{\beta\hbar\omega_D}{2\pi}\right)
			-\sum_{i=1}^3\ln\Gamma\left(\frac{\beta\hbar\lambda_i}{2\pi}\right) 
			-\ln \left(\frac{\beta\hbar\omega}{4\pi^2}\right),
\end{align}
where $\Gamma$ denotes the Euler Gamma function. In the absence of coupling, $\eta=0$, Eq.~\eqref{11} reduces to  $F(0, M_0) = 1/\beta \ln [2 \sinh( \beta \hbar \omega/2) ]$. In the low temperature, strongly damped limit, we find that $Q^{(C)}<0$, indicating that heat is dissipated into the environment:
\begin{align}\label{12}
 Q^{(C)}	&= - \frac{\hbar \omega b_0}{2 \pi} \, \left[1- \frac{\pi^2}{6 a^2}\right] \ .
\end{align}
The approximate, low temperature entropy change, $\Delta S^{(C)}= S(\eta, M_0) -  S(0,M_0)$, can be computed in a similar way as before and reads, 
\begin{align}\label{13}
\Delta S^{(C)} &=1+ \frac{1}{2} \left[ \ln \frac{2}{\pi^2} + \ln \ln c_0 + \ln \ln b_0 + \frac{\pi^2 b_0^2}{24 a^2 \ln b_0} \right].
\end{align}
It is worth noticing that, contrary to the case of the mass variation, Eq.~\eqref{9}, the entropy change induced by the  coupling to the reservoir is here positive, $\Delta S^{(C)}>0$.

{\it Clausius inequality.} We next consider the combined state transformation that consists of  the  initial coupling to the reservoir followed by the variation of the mass. We  accordingly define the total entropy change  $\Delta S =\Delta S^{(M)}+\Delta S^{(C)}$  and the total heat exchanged $Q = Q^{(M)}+ Q^{(C)}$. We again introduce a quantity $\Delta   =  Q - k T \Delta S$ which, using Eqs.~\eqref{8}, \eqref{9} and \eqref{12}, \eqref{13}, we can \mbox{write as}, 
\begin{align}
\label{14}
\Delta 		&=-kT - \frac{\hbar \omega b_1}{2 \pi} \, \times \\
			&\left[1 + \frac{\pi}{b_1 2a} \, \left[\ln \frac{2}{\pi^2} + \ln \ln c_1 + \ln \ln b_1\right] - \frac{\pi^2}{6 a^2}\right] \ .\notag
\end{align}

\begin{figure}[t]
\begin{center}
 \includegraphics[width=.48\textwidth]{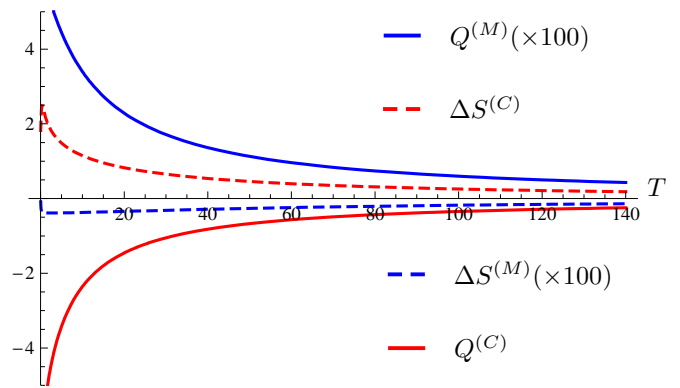} 
\end{center}
\caption{(Color online) Entropy and heat  during the initial coupling to the reservoir, $\Delta S^{(C)}$ (red dashed), $Q^{(C)}$ (red solid), and during a mass variation, $\Delta S^{(M)}$ (blue dashed), $Q^{(M)}$ (blue solid), as a function of temperature. Changes for the  initial coupling have opposite signs and much larger amplitudes than those for  the mass variation. Parameters are the same as in Fig.~\ref{f1}.} \label{f2}
\end{figure}
In the limit of large $a$, $b$ and $c$, this expression is always negative (see also Fig.~\ref{f1}). Thus, {\it no} violation of the Clausius inequality occurs in the low temperature, strong coupling regime, in agreement with standard thermodynamics. A better understanding  of the apparent violation found in Refs.~\cite{all00,all02,hor05,hil09,all01,hor08} can be obtained by comparing the change of entropy and heat during the two state transformations (see Fig.~\ref{f2}). The low-temperature  mass variation is characterized by a negative  entropy change and a positive heat, $\Delta S^{(M)}<0$, $Q^{(M)}>0$, which together lead to the breakdown of the Clausius inequality. On the other hand, the situation is exactly opposite for the initial reservoir coupling where $\Delta S^{(C)}>0$ and  $Q^{(C)}<0$. Figure \ref{f2} shows that in the limit of strong coupling, the latter contributions are much larger than those coming from the mass variation. They can therefore not be neglected as done so far in Refs.~\cite{all00,all02,hor05,hil09,all01,hor08}.

Deeper insight into the foregoing  discussion can be gained by using the concept of Hamiltonian of mean force \cite{rou99}. We express the reduced density operator of the system in the form $\rho_S= \exp[-\beta(H^*_S-F)]$, where
\begin{align} 
H^*_S= -\frac{1}{\beta} \ln \frac{\mbox{Tr}_B \exp(-\beta H)}{ \mbox{Tr}_B \exp(-\beta H_B)}
\end{align}
is the quantum Hamiltonian of mean force and $F= -1/\beta \,\ln \mbox{Tr}_S \exp(-\beta H^*_S)$ the free energy of the system. The quantity $\Delta H_S = H^*_S - H_S  $ vanishes for vanishing reservoir coupling and thus quantifies the deviation from a thermal state; it is simply related to the initial thermodynamic change of the system. We first note that the von Neumann entropy of the system is given by $S=\beta (U-F+\< \Delta H_S\>)$. The heat exchanged with the reservoir during the initial coupling is then $Q^{(C)}= kT \Delta S^{(C)} - \< \Delta H_S\>$ or, in other words, $\Delta^{(C)} = -\< \Delta H_S\>$. We therefore find that $\Delta = \Delta^{(M)} -\< \Delta H_S\>$. This result, valid for {\it any} quantum dissipative system, shows that the difference between the combined state transformation and the mass variation alone is just the difference between the Hamiltonian of mean force and the bare Hamiltonian of the system. It is worth stressing that for the quantum harmonic oscillator, $\<\Delta H_S\>$ is a function of the variances (2) and (3) of the reduced stationary state and can therefore be determined experimentally. 

The following physical picture thus  emerges from our analysis: In the limit of vanishing system-reservoir coupling, the stationary state of the system is thermal and the thermodynamic cost of the   coupling to the reservoir is negligible, $\Delta =\Delta^{(M)}$.  In the opposite limit of strong coupling, the stationary state becomes non-thermal, but the thermodynamic  contributions of the initial coupling  are important and hence need to be fully included, $\Delta = \Delta^{(M)} -\< \Delta H_S\>$.  In both cases, as we have just proved, the ordinary Clausius inequality $\Delta \leq0$ holds. We mention that an effective Clausius inequality has lately been derived  by introducing an effective  mass and spring constant for the oscillator, and an effective temperature, which  differs from that of the reservoir  \cite{kim10}; this approach also neglects the initial  reservoir coupling.

{\it Landauer's principle.} Let us finally derive the Landauer bound from the quantum Clausius inequality. We consider an isolated system with two stable states that are used to encode one bit of information (for instance a symmetric double-well potential with high energy barrier). The system is initially in equilibrium at temperature $T$ and the two states are occupied with equal probability. We reset the memory by first coupling it to the reservoir and then modulating the potential in order to bring the system with probability one into  one of its states \cite{ben82}. The von Neumann entropy of the system is hence $ \ln 2$ before the coupling to the reservoir and zero after complete erasure.   From the Clausius inequality, we then find that the dissipated heat obeys $Q_{dis} = - Q \geq -k T\Delta S = k T\ln 2$. This is Landauer's erasure principle.

{\it Conclusion.}
Our findings emphasize the crucial role of system-reservoir interactions in the thermodynamic description of quantum systems; a  low temperature  investigation  can therefore only be consistent if they are fully taken into account. We have derived detailed expressions for the change of entropy and heat induced by the coupling to the reservoir. We have further shown that their relative contributions grow with increasing interaction strength and even dominate in the strong coupling limit, thus safeguarding the validity of  Clausius' inequality. Contrary to previous claims that quantum correlations undermine Landauer's erasure principle, we have here demonstrated that the principle does hold when the generation of these correlations is properly included. Additionally, our results  provide a theoretical framework for the thermodynamic characterization of system-reservoir  correlations in quantum information theory \cite{hor02}.

JA thanks the Royal Society (London) for support in form of a Dorothy Hodgkin Fellowship. This work was further supported by the Emmy Noether Program of the DFG (Contract LU1382/1-1) and the
cluster of excellence Nanosystems Initiative Munich (NIM).

\end{document}